\documentclass[aps]{revtex4}

\def \beq {\begin{equation}}
\def \eeq {\end{equation}}

\begin{document}

\title{Non-Minimally Coupled Cosmology as  Geodesic Motion}

\author{Luciana A. Elias}
\author{Alberto Saa}
 \email{asaa@ime.unicamp.br}
  \affiliation{Departamento de Matem\'atica Aplicada, UNICAMP \\
   C.P. 6065,  13083-859 - Campinas, SP, Brazil.}


\begin{abstract}
Recent works showing that homogeneous and isotropic cosmologies involving 
scalar fields correspond to geodesics of   certain   augmented spaces are generalized to
the   non-minimal  coupling case. As  the Maupertuis-Jacobi principle in classical
mechanics, this result allows us, in principle, to infer some of the dynamical properties of the cosmologies
from the geometry of the associated augmented spaces.\\

\vspace{0.5cm}

{\footnotesize\sl \noindent Fifth International Conference on Mathematical Methods in Physics --- IC2006\\
		 April 24-28 2006\\
		 Centro Brasilerio de Pesquisas F\'\i sicas, Rio de Janeiro, Brazil}
\end{abstract}
\maketitle

\section{Introduction}

The Maupertuis-Jacobi principle\cite{Arnold} in classical mechanics establishes that the dynamics
of a given system can be viewed as geodesic motions in an associated Riemannian manifold. Recalling
briefly,  let us consider a classical mechanical
system with $N$ degrees of freedom described by the Lagrangian
\beq
\label{lagr}
{  L}(q,\dot{q}) = \frac{1}{2} g_{ij}(q)\dot{q}^i \dot{q}^j -
V(q),
\eeq
where $i,j=1, 2,\dots, N$,
the dot stands
for differentiation with respect to the time $t$, and
$g_{ij}$ is the Riemannian
metric on an $N$-dimensional configuration space $\mathcal M$. All the
quantities here are assumed to be smooth. The Euler-Lagrange
equations of (\ref{lagr}) can be written as
\beq
\label{EL}
\ddot{q}^i + \Gamma^i_{jk} \dot{q}^j \dot{q}^k = -
g^{ij}\partial_j V(q),
\eeq
where $\Gamma^i_{jk}$ is the Levi-Civita connection for the
metric $g_{ij}$.

The Hamiltonian of the system described by (\ref{lagr})
\beq
\label{ham}
{  H}(q,p) = \frac{1}{2} g^{ij}(q){p}_i {p}_j +
V(q),
\eeq
with $p_i=g_{ij}\dot{q}^j$,
is obviously a constant of motion, namely the total energy.
For a fixed energy $E$, the trajectories in the $2N$-dimensional phase-space $(q^i;p_j)$
are confined to the hypersurface
$E = \frac{1}{2} g^{ij}{p}^i {p}^j + V(q).$
On the other hand,
the admissible region for the trajectories in the configuration
space is given by
\beq
{\mathcal D}_E = \{q\in{\mathcal M} : V(q) \le E \}.
\eeq
In general, the region ${\mathcal D}_E$ can be bounded or not, connected
or not. The
boundary of the admissible region for the trajectories
is given by
\beq
\partial {\mathcal D}_E = \{q\in{\mathcal M} : V(q) = E \}.
\eeq
If the potential has no critical points on the boundary
$( V\ne 0)$, then
$\partial {\mathcal D}_E $ is a $N-1$ dimensional submanifold
of $\mathcal M$.
We can easily see that if a trajectory reaches the boundary
$\partial {\mathcal D}_E $ at a point $q_0$, its velocity at
this point vanishes and the trajectory approach
or depart from $q_0$ perpendicularly
to the boundary $\partial {\mathcal D}_E $. In particular,
there is no allowed trajectory along the boundary.

One can show that the equations of
motion (\ref{EL}) are, in the interior of
${\mathcal D}_E$,  fully equivalent to the geodesic equation
of the ``effective''
Riemannian geometry on $\mathcal M$ defined from the Jacobi
metric\cite{Arnold} 
\beq
\hat{g}_{ij}(q) = 2(E-V(q)) g_{ij}(q).
\eeq
The   geodesic
equation in question is given by
\beq
\label{geo}
\hat{\nabla}_u u =
\frac{d^2 q^i}{ds^2} + \hat{\Gamma}^i_{jk}
\frac{dq^j}{ds} \frac{dq^k}{ds} = 0,
\eeq
where
$u=dq^i/ds$ is the tangent vector along the geodesic and
$\hat{\nabla}$ and
$ \hat{\Gamma}^i_{jk}$ are, respectively, the covariant derivative
and the Levi-Civita connection
for the Jacobi metric $\hat{g}_{ij}$, and $s$ a parameter
along the geodesic obeying
\beq
\label{para}
\frac{ds}{dt} = 2 (E-V(q)).
\eeq
As any classical topic, there is a vast literature on the  Maupertuis-Jacobi principle.
We notice only that, motivated
by the celebrated result due to
Anosov\cite{avez} stating that the geodesic flow in a compact manifold with all
sectional curvatures negative at every point is chaotic, the Maupertuis-Jacobi principle
  has been 
recently invoked for the study of chaotic dynamics. (See, for instance,
\cite{chaos} and the references therein).

The main motivation of the present work is the the result presented in \cite{townsend}. The authors
considered cosmological models with $N$ self-interacting scalar fields $\phi^\alpha$ taking their values in
a Riemannian target space endowed with a metric $G_{\alpha\beta}$. The corresponding actions is
\beq
\label{act}
\int d^4x \sqrt{-g}\left(R - g^{ij}G_{\alpha\beta}(\phi)\partial_i\phi^\alpha \partial_j\phi^\beta - 2 V(\phi) \right).
\eeq
Greek indices run over $1\dots N$ (the target space dimension), while the lower case roman ones
run over $1\dots 4$ (the spacetime dimension). The spacetime metric is $g_{ij}$ and $R$ stands for
its scalar curvature. By considering the Friedman-Robertson-Walker homogeneous and isotropic metric
\beq
ds^2  = -dt^2 + a^2(t)d\Sigma_\kappa^2,
\eeq
$a(t)>0$, 
where $\Sigma_\kappa$ represents the 3-dimensional spatial sections of constant curvature $\kappa$,
they showed, by using arguments close to the Maupertuis-Jacobi principle, that the equations
of motion associated to the action (\ref{act}) do indeed correspond to the geodesics of a certain
``effective Jacobi" (pseudo) metric on an augment Lorentzian space. For the spatially flat
case ($\kappa=0$), for instance, the augmented space has $(1,N)$ signature and the geodesics
corresponding to the equations of motion derived from (\ref{act}) are   timelike, null,
or spacelike   according, respectively, if $V>0$, $V=0$, or $V<0$. These results have
been applied to the dynamical study of the models governed by actions of the type (\ref{act}), see \cite{appl}.

Applications of the
Maupertuis-Jacobi principle to the field equations obtained from   Hilbert-Einstein like actions
have also a long history. Non-homogeneous and anisotropic cases were considered in \cite{nh}.
Applications involving distinct differential
spaces instead of differential manifolds were discussed in
\cite{SHS}. Non-minimally coupled scalar fields, however, have not been considered so far.

\section{A Maupertuis-Jacobi principle for non-minimally coupled cosmology}

Non-minimally coupled scalar fields are quite common in cosmology. In particular, they have been invoked
recently to describe dark energy (see \cite{gunzig} and \cite{FS}, and the references therein, for,
respectively,   models using conformal coupling and   more general ones).
The non-minimal coupled generalization of (\ref{act}) we consider here
is
\beq
\label{act1}
\int d^4x \sqrt{-g}\left(F(\phi)R - g^{ij}G_{\alpha\beta}(\phi)\partial_i\phi^\alpha \partial_j\phi^\beta - 2 V(\phi) \right).
\eeq
In this work, we restrict ourselves to the $\kappa=0$ case, the full analysis will appear elsewhere\cite{LuSaa}.
For such a case, one has $R=6\dot{H}+12H^2$, $H=\dot{a}/a$. Integrating by parts the action (\ref{act1}), we
obtain the following Lagrangian 
\beq
\label{l1}
L(a,\dot{a},\phi_\alpha,\dot{\phi}^\alpha) = a^3(- 6 H^2 F - 6H  \dot{\phi}^\alpha\partial_\alpha F
+ G_{\alpha\beta}(\phi)\dot{\phi}^\alpha \dot{\phi}^\beta - 2V(\phi))
\eeq
for the system on the $N+1$ configuration space spanned by $(a,\phi_\alpha)$. By introducing the following Lorentzian metric
\beq
\label{m}
G_{AB}(a,\phi^\alpha) = \left( 
\begin{tabular}{c|c}
$-6aF $ & $-3a^2\partial_\beta F $ \\ 
\hline
$-3a^2\partial_\alpha F  $ & $a^3G_{\alpha\beta}$
\end{tabular} \right)
\eeq
on the configuration space (upper case roman indices run over $0\dots N$), the Lagrangian (\ref{l1}) can be 
cast in the form
\beq
\label{l2}
L(\phi_A,\dot{\phi}^A) = G_{AB}(\phi)\dot{\phi}^A \dot{\phi}^B - 2V_{\rm eff}(\phi),
\eeq
where $\phi^A= (a,\phi^\alpha)$ and $V_{\rm eff}(\phi^A) = a^3V(\phi^\alpha)$. It is clear the similarity
with (\ref{lagr}), provided that $\det G_{AB} \ne 0$, which we assume by now. We will return to  this issue 
in the last Section. 

Before considering the Maupertuis-Jacobi principle, let us recall that our manipulations imply that
all solutions of the Euler-Lagrange of (\ref{act1}) are also solutions of the Euler-Lagrange equations
of (\ref{l1}), but not the converse. Einstein equations form a constrained system. The solutions
of (\ref{act1}) correspond, indeed, to a subset of the solutions of (\ref{l1}), as one can realize by
considering the  Hamiltonian associated to (\ref{l1})
\beq
H(\phi_A, \pi_A) = G^{AB}\pi_A \pi_B + 2V_{\rm eff}(\phi) = a^3(- 6 H^2 F - 6H  \dot{\phi}^\alpha\partial_\alpha F
+  G_{\alpha\beta}(\phi)\dot{\phi}^\alpha \dot{\phi}^\beta + 2V(\phi)),
\eeq
which, obviously, must be a constant of motion, say $H(\phi_A, \pi_A) = E$. The Euler-Lagrange equations of
(\ref{act1}), on the other hand, implies that $E=0$ (the so-called energy constraint). Hence, we must
bear in mind that the relevant solutions of our original problem correspond indeed to the $E=0$ subset of the
dynamics governed by (\ref{l1}). Now, let us consider separately the cases $V=0$, $V<0$ and $V>0$. 


According to (\ref{EL}), for $V=0$ the Euler-Lagrange equations of (\ref{act1}) are already given
by geodesics of (\ref{m}). From the energy constraint
\beq
\label{ener}
 G_{AB}(\phi)\dot{\phi}^A \dot{\phi}^B = - 2a^3V(\phi),
\eeq
one sees that such geodesics are of null type.
For $V<0$, one can introduce the metric
\beq
\hat{G}_{AB} = -2VG_{AB}
\eeq
and obtain the same conclusions by repeating the same procedure used 
  for the classical Maupertuis-Jacobi principle. From the energy
constraint (\ref{ener}), it is clear that the associated geodesics will be spacelike. 
Finally, for $V>0$, one can introduce 
\beq
\hat{G}_{AB} =  2VG_{AB}
\eeq
and repeat exactly the same steps followed for the last case. In this case, the geodesics will be timelike.
Since the metric (\ref{m}) is not positive defined, one cannot, in general, obtain from the energy constraint a 
dynamically admissible region $\mathcal D$ of the configuration space, as done for  the classical Maupertuis-Jacobi
principle. Also as a consequence of the Lorentzian signature of (\ref{m}), one has that, typically,  $\mathcal D$
will be unbounded.

Summarizing, the equations governing the cosmological model (\ref{act1}) correspond to the geodesics of the
metric
\beq
\label{m2}
\hat{G}_{AB} = \left\{ \begin{array}{l}  G_{AB}\quad {\rm if}\ V=0, \\ 2|V|G_{AB}\quad {\rm if}\ V>0\ {\rm or}\ V<0,
\end{array} \right.
\eeq
with $G_{AB}$ is given by (\ref{m}). Moreover, since
\beq
\hat{G}_{AB}(\phi)\dot{\phi}^A \dot{\phi}^B = -4a^3|V|V,
\eeq
one has that the geodesics are of null type, timelike or spacelike, respectively, according if
$V=0$, $V>0$, or $V<0$. All the results\cite{townsend} that have motivated this work can be obtained by
setting $F=1$.

\section{Final Remarks}

Since the equations for cosmological models like (\ref{act1}) correspond to the geodesic
equations of the Lorentzian metric (\ref{m2}), it is possible, in principle, to obtain
some information about the cosmological dynamics from the geometry of the associated 
Lorentzian metric. For instance, it is shown in \cite{townsend} that cosmological
solutions exhibiting late time acceleration are related to geodesics entering in a certain
region corresponding to a subset of the   lightcone of (\ref{m2}). The study of dynamical singularities
can also benefit from these results. The essential assumption of $\det G_{AB} \ne 0$ is strictly
related to the avoidance of some singularities. For the $N=1$ and $G_{\alpha\beta}=1$ case, one
has
\beq
\det G_{AB} = -6a^4\left(F(\phi) + \frac{3}{2}\left( F'(\phi) \right)^2 \right).
\eeq
The vanishing of this quantity is known to be associated with the existence of some
unavoidable dynamical singularities (see, for references, \cite{sing}), which render the associated cosmological 
model unphysical.
Some preliminary results suggest that the same occurs for the $N$-field case.

We finish by noticing that one of the most
interesting  
peculiarities  of the conformal coupling $(F(\phi)=1-\phi^2/6)$ is that it can evade the $\det G_{AB} = 0$
singularity, since  
\beq
F(\phi) + \frac{3}{2}\left( F'(\phi) \right)^2 = 1
\eeq
for the conformal coupling  
and, consequently, the Maupertuis-Jacobi principle can be always employed.

 \acknowledgments
 The authors thank Dr. Ricardo Mosna for valuable help, and FAPESP and CNPq for the financial
 support.

\end{document}